%% file: esem-industry-testability.tex
\documentclass[conference]{IEEEtran}

\IEEEoverridecommandlockouts
\IEEEpubid{\makebox[\columnwidth]{\textbf{Preprint -- ESEM 2019}\hfill} \hspace{\columnsep}\makebox[\columnwidth]{ }}

\usepackage{cite}
\usepackage{amsmath,amssymb,amsfonts}
\usepackage{algorithmic}
\usepackage{graphicx}
\usepackage{textcomp}
\usepackage{xcolor}

\def\BibTeX{{\rm B\kern-.05em{\sc i\kern-.025em b}\kern-.08em
    T\kern-.1667em\lower.7ex\hbox{E}\kern-.125emX}}

\usepackage{subfigure}
\input{preamble}

\newcommand{\boxit}[1]{\vspace{0.3cm}
\noindent
\fbox{
\begin{minipage}{24em}
\emph{#1} 
\end{minipage}
}
}

\newcommand{\EPOF}{\emph{XYZ}\xspace}
\newcommand{\Post}{\emph{ABC}\xspace}
\newcommand{\blind}[1]{\textbf{BLINDED}\xspace}
\newcommand\keyword[1]{[#1]}

\makeatletter                  
\def\mdseries@tt{m}      
\makeatother                   
\begin{document}

\title{Testability First!}

\author{
	\IEEEauthorblockN{Mohammad Ghafari}
    \IEEEauthorblockA{
    \textit{University of Bern}\\
    Bern, Switzerland \\
    mohammad.ghafari@inf.unibe.ch
    }
    \and
    \IEEEauthorblockN{Markus Eggiman}
    \IEEEauthorblockA{
    \textit{University of Bern}\\
    Bern, Switzerland \\
    markus.eggimann@students.unibe.ch
    }
    \and
    \IEEEauthorblockN{Oscar Nierstrasz}
    \IEEEauthorblockA{
    \textit{University of Bern}\\
    Bern, Switzerland \\
    oscar.nierstrasz@inf.unibe.ch
    }
}

%\author{
%	\IEEEauthorblockN{BLINDED}
%	\IEEEauthorblockA{
%		University of BLINDED\\
%  		BLINDED\\
%  		blind@blind.blind
%  	}
%}

\maketitle

\IEEEpubidadjcol

\begin{abstract}

\keyword{Background}
The pivotal role of testing in high-quality software production has driven a significant effort in evaluating and assessing testing practices.
\keyword{Aims}
We explore the state of testing in a large industrial project over an extended period.
\keyword{Method}
We study the interplay between bugs in the project and its test cases, and interview developers and stakeholders to uncover reasons underpinning our observations.

\keyword{Results}
We realized that testing is not well adopted, and that \emph{testability} (\ie \emph{ease of testing}) is low.
We found that developers tended to abandon writing tests when they assessed the effort to be high.
Frequent changes in requirements and pressure to add new features also hindered developers from writing tests.

\keyword{Conclusions}
%We believe that testability, like software architecture, cannot be an afterthought; it must be addressed throughout the software lifecycle, and supported by appropriate policies, processes, and tools.
Regardless of the debates on test first or later, we hypothesize that the underlying reasons for poor test quality are rooted in a lack of attention to testing early in the development of a software component, leading to poor testability of the component.
However, testability is usually overlooked in research that studies the impact of testing practices, and should be explicitly taken into account.

\end{abstract}

\begin{IEEEkeywords}

testability, test quality, testing practices

\end{IEEEkeywords}

%%%%%%%%%%%%%%%%%%%%%%%%%%%%%%%%%%%%%%%%%%%%%%%%%%%%%%%%%%%%
\section{Introduction}
\seclabel{section}

Software engineers believe that high quality of software testing will lead to high quality in the software product.
However, it is not so clear why the state of practice does not truly reflect this established wisdom, as software testing often suffers from low coverage and low effectiveness.
%
%The value of automated testing has been widely acknowledged both in academia and industry, yet it is not so clear whether industrial testing practices truly reflect this established wisdom.
% Is systematic software testing the norm in the software industry? What is the quality of existing tests, and how can that quality be measured?

Monitoring the activity of about 2\,500 developers over 2.5 years shows that half of developers do not test, most programming sessions end without any test execution, and software developers only spend a quarter of their time engineering tests, whereas they think they test half of the time~\cite{Beller:2019}.
Consequently, software systems are often released under-tested.
What's worse, there are contradictory views on whether coverage criteria have a positive effect on finding faults.
Some studies claim no correlation between unit test coverage and post-unit test defects~\cite{Gren:2017}, whereas a more recent study of open source projects shows that, on average, about 30\% of defective methods are covered by JUnit tests, and the number of methods executed by a JUnit test is strongly related to that test uncovering a defect~\cite{Petric:2018}.
In order to improve the effectiveness of test suites, researcher suggest to focus on the strength of test oracles along with code coverage~\cite{Schwartz:2018}.
In the past years, there has been increasing effort to assess and improve the fault-detection capability of test suites.
The state-of-the-art technique is mutation testing, which checks the ability of a test suite to reveal artificially produced defects~\cite{Papadakis:2019}.
Nevertheless, this technique consumes enormous computing resources, thus hindering its application in practice.
Researchers suggest to reduce the execution time of mutation testing, for example, by verifying the quality of the test cases for each individual method, instead of the overall test suite quality~\cite{Vercammen:2018}.

%%%%%%%%
%\rA{
%The introduction introduces the concept of "software testability" as the effort required to write tests. Unfortunately, the link between this concept and the rest of the paper is missing. How was the testability assessed? What it the connection between the insights reported by the developers and the actual testability of the software analyzed? Once again, this is likely a matter of rephrasing the introduction to clarify the goals of the paper.
%}
%\on{Perhaps in the introduction we can explain how we determined testability in practice as we did in section III.A. I.e., ease of testing as determined by the three criteria.}
%\rC{
%Consider adding a forma high-level research goal in the introduction to drive the research and to get readers on board of the topic.
%}
%\mg{We first asked ``What is the practice of testing in the project''. We observed that testing is not well adopted. Then we asked ``Why testing is not well adopted in practice'', and from \emph{developers' insights} we found that testability is low. How about a RQ like ``What testing practices do developers follow and why?''. What do u think?}
%\on{Could be ok. Alternatively a more specific question, like ``Do developers apply a discipline of thorough testing in practice to achieve high test coverage?'' ``If not, what are the reasons?''}
%%

There is general consensus among software engineers that to have high quality tests software needs to be testable in the first place.
However, to the best of our knowledge, the significance of \emph{testability} in the quality of test suites has not received enough attention.
There exist many definitions of software testability~\cite{Garousi:2019}.
In this paper we use it to mean \emph{``ease of testing''} in terms of effort needed to write tests that are both effective in revealing defects, and manageable in the face of changes.

We studied the interplay between bugs and test cases in a large industrial project in which a deliberate effort was made to improve testing quality partway through the project's lifetime.
We affirm that many factors can influence and characterize the testability of the project, whereas state-of-the-art work that compares the effectiveness of various testing strategies in multiple projects often neglects to take into account the possibly different degrees of testability and other testing-related characteristics of these projects.

We observed both relatively low coverage and low effectiveness of tests.
Many components were not tested because they were considered hard to test.
It was common for bugs to be detected with the help of manual testing rather than by automated tests.
Developers were reluctant to write tests for code that was subject to frequent changes.
We observed that commitment from management and team dedication were important prerequisites to instill a healthy software testing culture.
Finally, many useful tools and techniques have come from research efforts, yet they were often either not well-known, or high-quality implementations were not available, thus hampering their adoption in this company.

Digging deeper, we found that the root of the problem appeared to be that, when testing was not considered early in the development process,
this led to software designs that made it more difficult to test the software \emph{post hoc}.
In contrast, when a deliberate management decision was taken to improve test coverage, this
%indeed 
led to better software designs that were easier to test, and thus to improvements in software quality.

%\newpage
%
Regardless of existing debates on when to start automated testing in a software project, we hypothesize that establishment of clear testing policies applied throughout the entire development lifecycle, and early attention to testing of each component are crucial to ensure high testability of the code base, and thus high-quality tests.
We therefore see a need to conduct new studies with multiple projects that share the same testing practices, and put similar effort into testing.

%\rB{
%What I miss in the paper are clearly investigated (qualitatively or quantitatively) chains of evidence what actually impacts testability and how testability can be improved.}
%\mg{I think this is not in the scope of this paper, but can be a future work! What do u think?}
%\on{Well, we conclude the discussion section and the conclusion with a call for studies to prove this. Maybe we can push that message more in the introduction too, or give some hints how to do this.}

The rest of this paper is organized as follows.
We introduce the case study in \autoref{sec:study}.
We then present our key findings in \autoref{sec:results},
followed by a discussion in \autoref{sec:discussion}.
In \autoref{sec:related} we discuss related work,
and conclude the paper in \autoref{sec:conclusions}.

%%%%%%%%%%%%%%%%%%%%%%%%%%%%%%%%%%%%%%%%%%%%%%%%%%%%%%%%%%%%
\section{Case Study}
\seclabel{study}

The goal of this work was to learn about the state of testing in a large logistics company in Switzerland, which we will call \Post.
In particular, we investigated \emph{whether developers adopt a consistent testing practice, and if not, what are the reasons}.
In the following we discuss the reasons behind choosing this case study, and then present our methodology.

\subsection{Motivation}

We study a digital platform for sending and receiving physical as well as electronic mail.
We chose to study this platform for the following reasons:
(i) it is a large real-world software system comprising 14\,102 files with around half a million lines of Java and C\# code, and about 15\,500 issues,
(ii) we have access to five years of development history from 2013 to 2018, easing the recovery of links among commits and issues, and
(iii) we could approach over 40 developers and stakeholders for any matter about the project.

This platform, which we will call \EPOF, started as a simple web application in 2011 and grew over time.
The back-end is a large monolithic system that had its origins in that web application.
There were automated unit tests for testing the business logic, but the coverage was generally low.
As the project grew, it became increasingly difficult to manage.
In the summer of 2017, the team decided to structure the whole system in a more modular way.
New features should be added as modules, and these modules should be covered by unit tests.
The team and the project management agreed to strive for 40\%-80\% code coverage of those new modules.
Moreover, a team of testers would manually test the portal and the mobile apps, and manual regression tests would be executed after every development sprint.

\begin{figure}
    \centering
    \includegraphics[width=\linewidth]{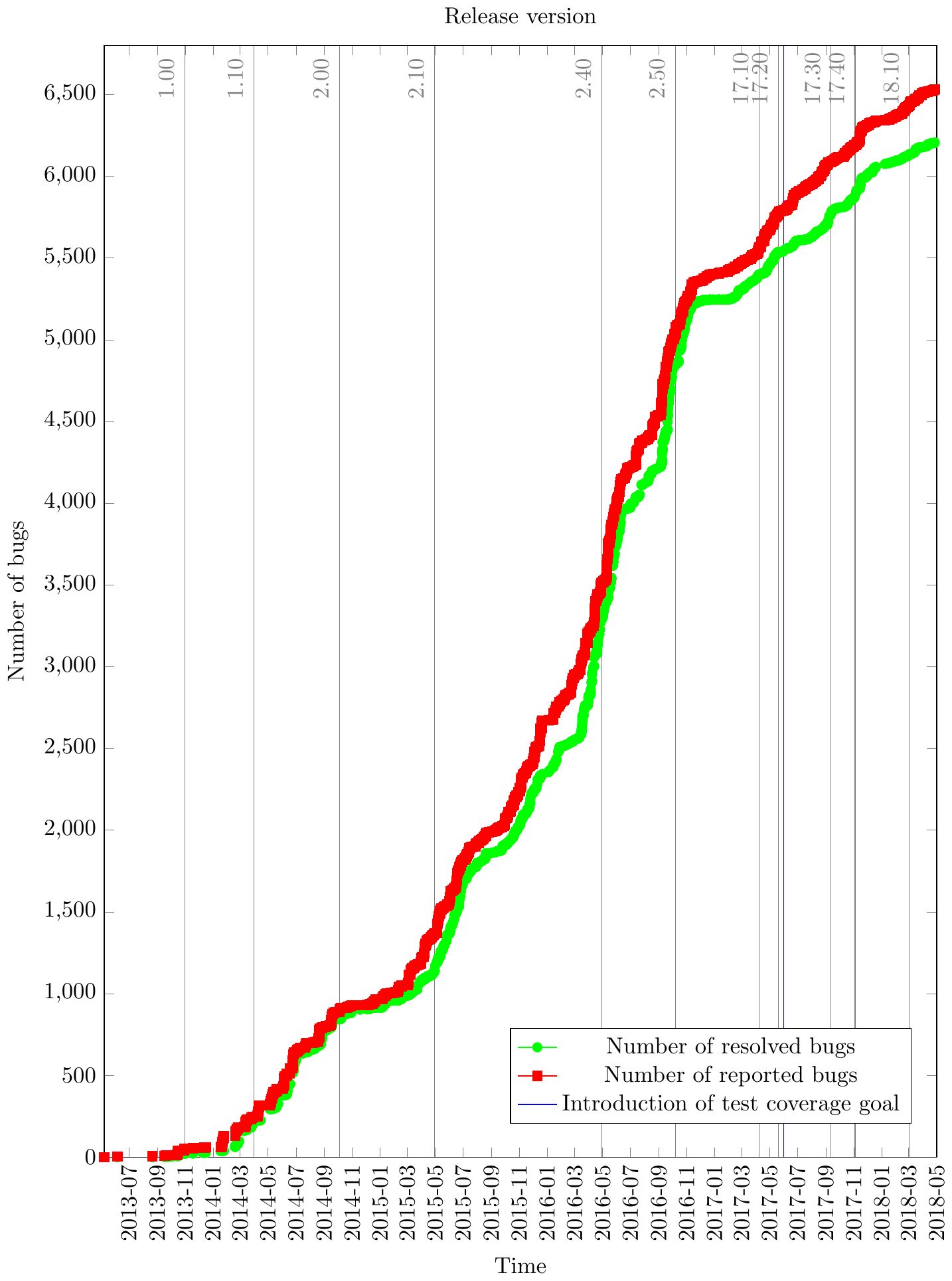}
    \caption{Bug discovery and fixes over time
    % Y axis should be ``Cumulative number of bugs'', and put circled letters (A), (B) to indicate the points of interest referred to in the text
    }
    \label{fig:discoverybugs}
\end{figure}

\autoref{fig:discoverybugs} shows the history of bugs in the \EPOF project.
The Y axis presents the cumulative number of bugs, and the X axis shows the time span of the project.
The red dots show the total number of reported bugs at a given point in time, and the green dots show the number of resolved bugs.
The gray vertical lines show the project releases.

% constant number of open bugs
Most of the time the green line follows closely behind the red line, which means that the total number of open bugs in the system at a given point in time remains more or less constant (on average 210 bugs).
Whereas developers were continuously fixing bugs throughout the lifetime of the project, the green line and the red line take different paths after November 2016, and the number of open bugs slightly increases to 249 bugs, on average.

Contrary to our expectation, we do not observe any decrease in the number of bugs with the introduction of testing policies and increased effort on testing in this project.
This motivated us to dig deeper into the state of testing in this company, and to develop  hypotheses for further study.
% In this work we shed light on reasons underpinning this phenomenon.

\subsection{Methodology}

% Bug report
In \EPOF every bug, whether identified by a developer or other stakeholders, will be reported in the Jira issue tracker with an issue  labeled as a ``bug".
Each bug report comprises a description of the bug, and the steps that others can follow to reproduce it.
% Bug fixing
Every project in the company is versioned on a Git repository, which is hosted on a Bitbucket server.
Developers usually create a new branch once they start to fix a bug, and should mention the issue key when committing a fix. 
% Recover links
Internally Jira is linked to Bitbucket server, and keeps track of branches and commits that are relevant to an issue. 
Jira provides a REST API through which we can recover this information. 

\newpage
We could automatically recover commits that fixed 1\,119 bugs. 
Of these bugs, 30\% (\ie 334 bugs) were reported after the testing effort increased in 2017.
For each bug we collected commits, file changes associated with those commits, and the developer information.

We were interested to understand the testing practices, and their impact on the quality of the \EPOF project. 
In particular, we investigated which components in the project had bugs, why these bugs existed, whether the components were tested before a bug report, and whether developers wrote tests during a bug fix.
As it is a non-trivial task to acquire this information for every bug in the project, we randomly chose 200 bugs, and manually inspected their associated commits.
Of the 200 bugs, 125 were reported after the introduction of the code coverage policy.

Together with a senior developer in the \Post company who was familiar with the \EPOF project, for every bug, we checked which components are involved.
% Markus at the time of the interview had 8 years of experience (6 years of professional programming experience. He was employed by \Post for 6 years, but he was never a member of \EPOF.
We then inspected whether the code part that caused the bug was covered by a unit test.
%%
%\rA{
%How did the authors practically analyze the reasons why tests did not fail? What was the process followed to come up with one or more reasons? Can the authors define a kind of taxonomy of causes leading tests to fail?
%}
%%
If so, we analyzed why these tests did not fail and, consequently, did not detect the bug.
We manually identified the so-called \emph{``focal methods under test''} to understand the actual purpose of each test case~\cite{Ghafari:2015}.
We also checked whether developers modified, added, or removed tests during the bug fix.

We interviewed the development team members and stakeholders regarding testing practices in the project.
We mainly interviewed eight senior developers who each had a minimum of eight years programming experience.
The interviews were conducted face-to-face within the company or on the phone, and they were open-ended.
In each interview, we usually explained relevant findings in the experiment first, and then asked a question. 
We encouraged the interviewees to freely discuss any details relevant to the topic.
% interview analysis
We transcribed the interviews, and analyzed them afterwards.

%%%%%%%%%%%%%%%%%%%%%%%%%%%%%%%%%%%%%%%%%%%%%%%%%%%%%%%%%%%%
\section{Results}
\seclabel{results}

In this section we present the observations that we drew from studying 200 bugs, followed by insights that developers and stakeholders shared with us in this regard.

% ==========================================================
\subsection{Observations}

\begin{figure}
    \centering
    \includegraphics[width=\linewidth]{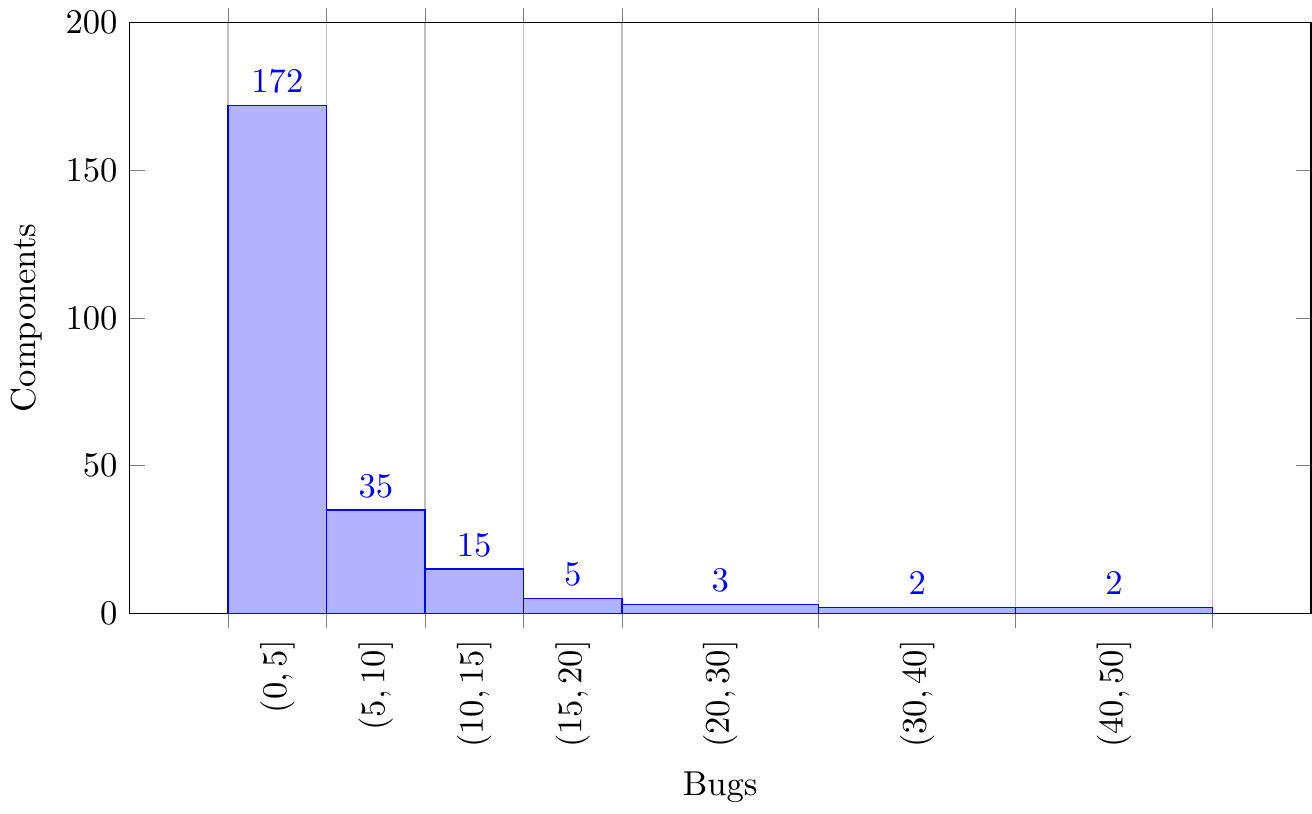}
    \caption{Bugs per component for $236$ untested components}
    \label{fig:bugfixPerComponentUntested}
\end{figure}

% tested components have fewer bugs
We analyzed two hundred randomly chosen bugs in depth.
% In 17 out of 200 analyzed bugs (~10\%), the defective component had associated unit tests.
Of the 244 components affected by those bugs, only eight components were covered by unit tests.
We then evaluated how many bugs affected those eight tested components and the remaining 236 untested ones.
Our investigation showed that all but one of the components under test were affected by five or fewer bugs; just one was involved in 19 bugs.
Further examinations revealed that this component handles the fingerprinting feature via a new Android API, and it took some time for the developers to learn how to use it in the right way.
In contrast to tested components, about a quarter of the untested components suffered from more than five bugs.
Furthermore, we could identify eight pairs of recurring bugs, or more precisely similar bugs in the project.
We also found a few tests that were commented out during the bug fix, and still commented in the latest version.
Figure \ref{fig:bugfixPerComponentUntested} illustrates the distribution of bugs in the untested components.
Tested components appear to be less prone to bugs than untested ones.

% bug fixes touched multiple files
For over half of the analyzed bugs, multiple files needed to be changed to fix them.
% classes = components = files 
Such cases might be hard to cover with simple unit tests, but might require integration tests.
%
% Bugs had different origins: source code, configuration files, libraries and platforms
We also discovered that some bugs are due not to coding errors but to changes in the underlying platform or to libraries.
Interestingly 20 bugs required no changes to program source file to fix them, but rather to other kinds of files like configuration or resource files that were excluded from the analysis.

% testability level
%We (the first two authors) 
We manually inspected the code associated with each of the 200 randomly chosen bugs, and identified 140 cases that we considered ``hard to test'' based on project domain knowledge.
We identified three main reasons:
(i) a component violates the single responsibility principle,
(ii) it has many dependencies on other parts of the system or external components, or
(iii) no exact definitions of correct or incorrect behavior exist, \eg a progress bar appearing in a wrong location.
We observed that developers almost never wrote tests for such cases.
In only three cases, \ie 2\%, the defective components were partly covered by tests.

There were only 16 cases that we did not consider hard to test.
In seven of those, there were indeed tests that at least partly covered the defective component.
Some of those tests existed before the bug was discovered, and some were added when the bug was fixed.

We then asked ourselves what is the quality of the existing tests.
% Mutations -- low quality
%\rC{
%Did I understood correctly that the authors introduced additional tests? and or mutation testing for existing tests? for selected components?
%}
%\mg{I'm not sure where is the issue?}
%\on{I think the reviewer is confused by the mutation testing. He thinks we added new tests to the project. But the mutation tests are to test the quality of the existing tests. I'm not sure how to address the conufsion, as the text seems clear to me.}
We consequently applied mutation testing to assess whether the existing tests could detect changes to the source code that were covered by the tests.
Our analysis showed that the existing tests are generally not good at detecting code manipulations, however, tests that were introduced during the release 17.XX onward obtained a significantly higher mutation score in our experiment than older tests.
Curiously, none of the new components were heavily modified, though according to the bug evolution (\autoref{fig:discoverybugs}) the average number of open bugs increased.
To understand the reason for this, we approached the developers.

\boxit{
Tested components had fewer bugs than other components. 
However, a majority of components were hard to test, thus impacting test quality.
}

% ==========================================================
\subsection{Developers' Views}

% We never report observations based on a single case/opinion, and present only what the development team unanimously agreed on.

We approached team managers and developers who were mainly involved in the top 15 of the most error-prone components, and interviewed them about their testing practices.

We asked developers \emph{``whether writing tests is important, and if so why the test coverage is so low.''}
Developers acknowledged the high value of testing, but they also considered it less important (and perhaps less glamorous) than other tasks, such as developing new features.
They stated that over roughly a two-year period, the team was under constant pressure to add more features.
Furthermore, the communication between the customer and the development team was not very effective.
The team had already started implementing new features before all requirements were clear.
This led to misunderstandings between the developers and the customers and resulted in many ``bug reports'', which in fact were changes in requirements.
Developers stated that the high frequency of changes demotivated them from writing tests.
The team also suffered from high turnover.
As a consequence, the project team neglected testing and quality assurance activities, which resulted in many bugs that ended up in production.

% improved quality due to deliberate policy
The team manager explained that after December 2016, the team decided to improve its Scrum process and follow it more strictly.
They improved the quality of communication between the customer and the development team to be more open and better structured.
Quality assurance, especially automated unit testing, was given higher priority, and a test coverage goal was introduced and monitored using the SonarQube tool.\footnote{http://www.sonarqube.org}
As a consequence, the quality of the product improved, which is confirmed by feedback of both customers and developers, as well as by a significantly higher mutation score in our experiment.
%
% why bugs became more
We next queried him regarding the slight increase in the average number of open bugs starting in that period (see \autoref{fig:discoverybugs}).
He stated that this is a consequence of a ``cleanup'' measure in which developers (rather than customers) started to systematically report annoying but often non-critical issues, which apparently led to the slight average increase in open bugs.
Therefore, the increase in the number of bug reports did not necessarily indicate a decrease in the product quality.

\boxit{Developers were aware of the value of testing, however, frequent changes in the requirements, and pressure to add new features hindered developers from writing tests.
Improved communication between stakeholders, and a clear test coverage policy improved the quality of the product.
}

In most cases bugs were fixed without making changes to any tests.
We asked developers \emph{``when do you write tests?''}.
We found that when the effort for developing a unit test for a component appeared to be high, developers tended to abandon writing tests for those components.
Developers confirmed that several components are rather large and play an important role in the application are prone to bugs, but are also very hard to test.
For instance, some key components that were hard to test already at the start of the project, grew and became more complex over time.
Team members said that they never anticipated the importance of these components in the actual product, and so their testability was not considered at the time.
In the experiment, we found three defective components were partly covered by tests, though they were hard to test.
Unfortunately, developers of these test cases were no longer available to be asked what motivated them to develop such hard-to-write tests.

\boxit{Developers tended to give up writing tests for a component that is hard to test.}

We asked \emph{``does the discovery of bugs lead to new tests being written''}, and realized that developers do not follow the practice of writing new tests that reproduce reported bugs, though they knew it.
We found that developers generally ignore this advice especially when they are confident that they can easily reproduce the bugs manually. 
They usually fix what they assume to be trivial bugs without writing tests.
Of the eight pairs of recurring bugs, all were detected by testers who performed manual tests before rolling out the software.
It seemed that none of the developers were aware of recurring bugs, and unanimously agreed that they would have written a test if they had been aware of the recurrence of these bugs.
Further discussions with developers revealed that they seldom consider the impact of a bug (\ie its severity) to decide whether or not to write tests, which is staggering.
They mentioned that if a bug is critical, it is fixed immediately and the corrections are deployed as a hotfix. Otherwise, it is moved to the backlog of the project and resolved in one of the next development sprints.
We noticed that developers never had a formal training for testing, and tests were only reviewed in an ad hoc way, without the help of specific guidelines.

\boxit{Developers decided to write a test mostly based on their personal opinion rather than upon a clear guideline. The impact of a bug barely influenced this decision.}

Finally, when we asked \emph{``what specific tools do you use to facilitate testing"}, 
developers only named the SonarQube tool. 
We then particularly asked if they use any tool, for instance, to (i) recover the traceability links between tests and production code, (ii) assess the quality of tests, and not just their coverage, and (iii) recognize recurring bugs.
The answer was mainly negative: few developers knew about a couple of free mutation testing tools, but they said that they crash and do not work as they expected.

% ==========================================================
\subsection{Threats to Validity}

We have drawn our observations based on a single case, which is reflective of other similarly-sized software projects in the \Post company. However, the results may not generalize to the state of practice in other companies in Switzerland or abroad.

The findings of this experiment were limited to a subset of randomly chosen bugs in the \EPOF project that we could automatically link with commits.   
We manually studied the bugs, and associated commits and test cases, which could have been imprecise due to a lack of experience with the project. 
We reduced this threat by asking a senior developer in the \Post company to help us throughout the whole journey.
Also, whenever an issue arose that we could not resolve together, we approached team members and stakeholders of the \EPOF project.

We did our best to interview the right person for each matter.
For instance, to learn about recurring bugs, we asked developers who fixed such bugs.
Nevertheless, in cases where the right person was not available we queried a peer in the development team or the manager.
Moreover, to reduce bias during the interviews, we kept our questions open-ended, and encouraged the interviewees to discuss any matter related to the topic.
The interviews were conducted in German and English, and it may be possible that we misunderstood the answers. 
We mitigated this threat by double-checking our conclusions with the practitioners.

%%%%%%%%%%%%%%%%%%%%%%%%%%%%%%%%%%%%%%%%%%%%%%%%%%%%%%%%%%%%
\section{Discussion}
\seclabel{discussion}

%\rA{
%The last (big) problem of the paper concerns the implications of the study. Looking at the findings, a reader could somehow guess what would be the practical actionable items; however, they should clearly be reported in the paper. Section IV does a preliminary work in this sense, but I would recommend the addition of a more structured organization in which the authors could specify (i) what are the implications of the study and (ii) how these can be tackled by researchers and/or practitioners.
%}
%\mg{I have no clue what to do. Any recommendation?}
%\on{I think the last two paragraphs are rather fuzzy. He wants us to provide a list of concrete actions that could be taken. We could give a hint of the kinds of policies that should be put into place, how to assess testability, how to ensure early testability, how to assess the impact of the policies and i,prove them ...}

We observed that testing is limited in the \EPOF project, and from developers' insights we found that the root cause of this phenomenon is that the testability of the project is low.

We see a clear need to encourage developers to start testing early, mainly to ensure testability \ie the possibility to write automated tests without requiring high effort.
We believe realizing the ``testability" goal requires commitment from both management and from the development team.
For instance, management should establish
regular training about software testing, 
guidelines for writing and reviewing tests,
and recognition for testing efforts as being equivalent in value to development tasks.
Moreover, developers should commit to a culture of testing.
Finally, tool support may exist either as research prototypes or as industrial products, but only few tools are adopted in practice; both industrial applicability and awareness of these tools must be improved.

There are enduring debates in the state-of-the-art work that studies the impact of testing practices.
For instance,
%%%%%%
Bissi~\etal conduct a systematic review to identify publications that compare the effect of 
Test Driven Development (TDD) and Test Later Development (TLD)~\cite{Bissi:2016}.
They conclude from the findings of previous studies that TDD yields more benefits than TLD for internal and external software quality, but it results in lower developer productivity.
%%%%%%
Pancur~\etal compare TDD and TLD based on productivity, code coverage, fault-finding capabilities, \etc~\cite{Pancur:2011}.
The experiment is conducted with fourth-year undergrad students during a course on distributed systems.
The students are introduced in general to testing as well as TDD and TLD.
They conclude that the benefits of test-driven development compared to iterative test-last development are small, and there is no difference regarding productivity.
%% Issues
%Students may have various experiences and goals when attending a course.
%They may not be proper representative of professional population.
%Even worse, professionals in one environment may not be representative of professionals in other environments, even within the same application domain.
%%%%%%
Fucci \etal conduct an experiment with professionals in two companies.
From four runs of a workshop about unit testing and TDD at these two companies, they find that the order in which test and production code are written has no important influence on software quality or developer productivity~\cite{Fucci:2017}.
They conclude that the claimed benefits of TDD may be due to the fact that TDD-like processes encourage fine-grained, steady steps that improve focus and flow.
%%%%%%
Finally, Borle \etal study open source Java projects that have adopted TDD to some extent.
They find that this practice is relatively rare in GitHub projects, and that the characteristics of projects with and without TDD are almost the same~\cite{Borle:2018}.
These studies were conducted in different and incomparable settings, for example one with a student project, and the other with professionals through a course of workshops.
The subjects had different levels of experience, and the effort they put into testing is likely different.
Consequently, the level of testability in these projects may vary, and so we cannot easily draw any valid conclusions about the relative benefits of the different practices.

Regardless of the contradictory views on test first or later, we argue the need for \emph{``testability first!''}: 
we hypothesize that establishment of clear testing policies applied throughout the entire development lifecycle, and early attention to testing of each component are crucial to ensure high testability of the code base, and thus high-quality tests.

In order to draw reliable conclusions about the impact of testing practices, such as whether writing tests before code is advantageous,
we therefore see a need to conduct new studies with multiple projects that share the same level of testability: similar effort has been put into testing from management to the development team.

%%%%%%%%%%%%%%%%%%%%%%%%%%%%%%%%%%%%%%%%%%%%%%%%%%%%%%%%%%%%
\section{Related work}
\seclabel{related}

% Garousi \etal present a systematic study of advice on what to test and when to write automated tests~\cite{Garousi:2016}. They enumerate a few factors such as software-related factors, testing tool-related factors, and human and organizational factors that are all strongly context dependent.
%
%Garousi \etal conduct a comprehensive study of smells in test code, and provide a large catalogue of test smells, along with a summary of guidelines, techniques and tools to deal with such smells~\cite{Garousi:2018}.

%Beller \etal study how developers test in real-world projects by monitoring the activity of about 2\,500 developers over 2.5 years~\cite{Beller:2019}.
%They report that half of developers do not test, most programming sessions end without any test execution, and software developers only spend a quarter of their time engineering tests, whereas they think they test half of the time.

Kochhar~\etal interview and survey practitioners to understand the characteristics of good test cases and testing practice~\cite{Kochhar:2019}.
For instance, practitioners agree that high coverage does not mean that a test suite can detect more bugs, and that designing tests that cover different requirements is often superior to maximizing code coverage.
They also strongly concur with writing a test that cover a bug fix.

Petric \etal study how effectively defective code is actually tested in seven open source Java projects, and show that most of these projects are under-tested.
On average about 30\% of defective methods are covered by JUnit tests, and the number of methods touched by a JUnit test is strongly related to that test uncovering a defect~\cite{Petric:2018}.

Gren and Antinyan study the relationship between unit testing and code quality, and find no correlation between unit test coverage and post-unit test defects~\cite{Gren:2017}.

Schwartz \etal investigate the cause behind contradictory answers to the question whether coverage criteria have a positive effect on finding faults~\cite{Schwartz:2018}. 
They find that test suites with high code coverage are not able to find specific types of faults as frequent as other types.
They conclude that code coverage alone does not ensure that faults are triggered and detected, and that the selection of input and oracle can improve the effectiveness of test suites.

Toure~\etal investigate the ability of a synthetic metric called ``Quality Assurance Indicator'' to predict early the testing effort in object-oriented software systems~\cite{Toure2018}.
They analyze eight open-source Java projects, and show that the models trained based on this metric have a promising performance to predict the effort required for writing unit test cases.

% Beer, A. et al.: Measuring and improving testability of system requirements in an industrial context by applying the goal question metric approach. RET@ICSE 2018: 25-32.

Garousi~\etal summarize a pool of 208 papers that help both practitioners and researchers to prepare, measure, and improve software testability~\cite{Garousi:2019}.
They report that observability and controllability are the two most frequent factors affecting testability, and common ways to improve testability are testability transformation, improving observability, adding assertions, and improving controllability.

%%%%%%%%%%%%%%%%%%%%%%%%%%%%%%%%%%%%%%%%%%%%%%%%%%%%%%
\section{Conclusions}
\seclabel{conclusions}

We have presented our observations on the state of testing in a large industrial software project.
We affirmed that various factors can influence and characterize the testability of a project.

% What is the quality of tests in practice?
We found that test coverage was relatively low until a deliberate policy was put in place to improve it.
Mutation testing revealed that existing tests were often of low quality.

%What motivates developers to write test cases?
In general developers gave up writing automated tests when the effort was high, though there was incentive to write tests for critical components.
When they were pressured to prioritize development of new features over tests, test coverage suffered.
A deliberate management policy to improve test coverage led to an increase in tests and in the quality of the code.

% When do developers write tests?
Developers were reluctant to test software that undergoes a high rate of change.
They typically did not write new tests while fixing bugs, and the bug severity had no impact on writing tests.
Developers were not aware of recurring bugs,  and therefore such bugs were mostly captured during manual testing.

We observe a lack of attention to the testability of software in the state-of-the-art work that studies the impact of testing practices and methodologies.
We hypothesize that establishment of clear testing policies applied throughout the entire development lifecycle, and early attention to testing of each component are crucial to ensure high testability of the code base, and thus high-quality tests.
We therefore see a need to conduct new studies with multiple projects that are similar in terms of testability.

%Comparison on the perception of testing in practices versus academia
%Investigating (qualitatively or quantitatively) chains of evidence what actually impacts testability and how testability can be improved.
%How do we measure testability?
%What metrics, attributes, smells etc can reliably indicate problems with testability?
%What practices can improve testability, and how to we validate that they work?

%%%%%%%%%%%%%%%%%%%%%%%%%%%%%%%%%%%%%%%%%%%%%%%%%%%%%

% \newpage

\section*{Acknowledgment}
We appreciate the entire development team and stakeholders of the \EPOF project who supported us during this experiment.
We also thank the anonymous reviewers for their valuable feedback that led to a more coherent version of this paper.

We gratefully acknowledge the financial support of the 
Swiss National Science Foundation for the project ``Agile Software Assistance'' (SNSF project No.\,200020-181973, Feb.\,1, 2019 - April 30, 2022).

%%%%%%%%%%%%%%%%%%%%%%%%%%%%%%%%%%%%%%%%%%%%%%%%%%%%%

\bibliographystyle{plain}
\bibliography{bibliography}

\end{document}

%% file: preamble.tex
\usepackage{xspace}
\usepackage{graphicx}
\usepackage{ifthen}
\usepackage[normalem]{ulem} % for \sout
\usepackage{xcolor,colortbl}

\usepackage[bookmarks=false]{hyperref}
\newboolean{showedits}
\setboolean{showedits}{true} % toggle to show or hide edits
\ifthenelse{\boolean{showedits}}
{
	\newcommand{\del}[1]{\textcolor{red}{\sout{#1}}} % please delete
	\newcommand{\nbe}[3]{
		{\colorbox{#3}{\bfseries\sffamily\scriptsize\textcolor{white}{#1}}}
		{\textcolor{#3}{\sf\small$\blacktriangleright$\textit{#2}$\blacktriangleleft$}}}
}{
	\newcommand{\del}[1]{} % please delete
	
	\newcommand{\nbe}[3]{}
}

\newboolean{showcomments}
\setboolean{showcomments}{true} 
\newcommand{\id}[1]{$-$Id: scgPaper.tex 32478 2010-04-29 09:11:32Z oscar $-$}

\ifthenelse{\boolean{showcomments}}
 {
 	\newcommand{\nbc}[3]{
 		{\colorbox{#3}{\bfseries\sffamily\scriptsize\textcolor{white}{#1}}}
		{\textcolor{#3}{\sf\small$\blacktriangleright$\textit{#2}$\blacktriangleleft$}}}
	
 }{
 	\newcommand{\nbc}[3]{}
 	
 }

\usepackage[most]{tcolorbox}
\ifthenelse{\boolean{showedits}}
{
  \newtcolorbox{inserted}{%
       title=Inserted text:,
       colframe=blue,colback=blue!5!white,
       breakable,
       leftrule=0mm, 
       bottomrule=0mm,
       rightrule=0mm,
       toprule=0mm,
       arc=0mm, outer arc=0mm,
       oversize
  }
  \newtcolorbox{deleted}{%
       title=Deleted text:,
       colframe=red,colback=red!5!white,
       breakable,
       leftrule=0mm, 
       bottomrule=0mm,
       rightrule=0mm,
       toprule=0mm,
       arc=0mm, outer arc=0mm,
       oversize
  }
  \newtcolorbox{refactored}{%
       title=Rewritten text:,
       colframe=blue,colback=red!5!white,
       breakable,
       leftrule=0mm, 
       bottomrule=0mm,
       rightrule=0mm,
       toprule=0mm,
       arc=0mm, outer arc=0mm,
       oversize
  }
}{

}

\newcommand{\seclabel}[1]{\label{sec:#1}}

\newcommand{\commented}[1]{}
\newcommand{\eg}{\emph{e.g.,}\xspace}
\newcommand{\ie}{\emph{i.e.,}\xspace}
\newcommand{\etal}{\emph{et al.}\xspace}
\newcommand{\etc}{\emph{etc.}\xspace}

%% file: esem-industry-testability.bbl
\begin{thebibliography}{10}

\bibitem{Beller:2019}
Moritz Beller, Georgios Gousios, Annibale Panichella, Sebastian Proksch, Sven
  Amann, and Andy Zaidman.
\newblock Developer testing in the {IDE}: Patterns, beliefs, and behavior.
\newblock {\em IEEE Transactions on Software Engineering}, 45(3):261--284,
  March 2019.

\bibitem{Bissi:2016}
Wilson Bissi, Adolfo Gustavo Serra~Seca Neto, and Maria Claudia
  Figueiredo~Pereira Emer.
\newblock The effects of test driven development on internal quality, external
  quality and productivity: A systematic review.
\newblock {\em Information and Software Technology}, 74:45 -- 54, 2016.

\bibitem{Borle:2018}
Neil~C. Borle, Meysam Feghhi, Eleni Stroulia, Russell Greiner, and Abram
  Hindle.
\newblock Analyzing the effects of test driven development in {GitHub}.
\newblock {\em Empirical Software Engineering}, 23(4):1931--1958, Aug 2018.

\bibitem{Fucci:2017}
Davide Fucci, Hakan Erdogmus, Burak Turhan, Markku Oivo, and Natalia Juristo.
\newblock A dissection of the test-driven development process: Does it really
  matter to test-first or to test-last?
\newblock {\em IEEE Transactions on Software Engineering}, 43(7):597--614, July
  2017.

\bibitem{Garousi:2019}
Vahid Garousi, Michael Felderer, and Feyza~Nur Kılı{\c c}aslan.
\newblock A survey on software testability.
\newblock {\em Information and Software Technology}, 108:35 -- 64, 2019.

\bibitem{Ghafari:2015}
Mohammad Ghafari, Carlo Ghezzi, and Konstantino Rubinov.
\newblock Automatically identifying focal methods under test in unit test
  cases.
\newblock In {\em 2015 IEEE 15th International Working Conference on Source
  Code Analysis and Manipulation (SCAM)}, pages 61--70, Sep. 2015.

\bibitem{Gren:2017}
Lucas Gren and Vard Antinyan.
\newblock On the relation between unit testing and code quality.
\newblock In {\em 2017 43rd Euromicro Conference on Software Engineering and
  Advanced Applications (SEAA)}, pages 52--56, Aug 2017.

\bibitem{Pancur:2011}
Matjaž Pančur and Mojca Ciglarič.
\newblock Impact of test-driven development on productivity, code and tests: A
  controlled experiment.
\newblock {\em Information and Software Technology}, 53(6):557 -- 573, 2011.
\newblock Special Section: Best papers from the APSEC.

\bibitem{Papadakis:2019}
Mike Papadakis, Marinos Kintis, Jie Zhang, Yue Jia, Yves~Le Traon, and Mark
  Harman.
\newblock Chapter six --- {Mutation} testing advances: An analysis and survey.
\newblock volume 112 of {\em Advances in Computers}, pages 275 -- 378.
  Elsevier, 2019.

\bibitem{Kochhar:2019}
Kochhar Pavneet~Singh, Xin Xia, and David Lo.
\newblock Practitioners' views on good software testing practices.
\newblock In {\em Proceedings of the 41st International Conference on Software
  Engineering}, ICSE '19, 2019.

\bibitem{Petric:2018}
Jean Petri\'{c}, Tracy Hall, and David Bowes.
\newblock How effectively is defective code actually tested?: An analysis of
  {JUnit} tests in seven open source systems.
\newblock In {\em Proceedings of the 14th International Conference on
  Predictive Models and Data Analytics in Software Engineering}, PROMISE'18,
  pages 42--51, New York, NY, USA, 2018. ACM.

\bibitem{Schwartz:2018}
Amanda Schwartz, Daniel Puckett, Ying Meng, and Gregory Gay.
\newblock Investigating faults missed by test suites achieving high code
  coverage.
\newblock {\em Journal of Systems and Software}, 144:106 -- 120, 2018.

\bibitem{Toure2018}
Fadel Toure, Mourad Badri, and Luc Lamontagne.
\newblock Predicting different levels of the unit testing effort of classes
  using source code metrics: a multiple case study on open-source software.
\newblock {\em Innovations in Systems and Software Engineering}, 14(1):15--46,
  Mar 2018.

\bibitem{Vercammen:2018}
Sten Vercammen, Mohammad Ghafari, Serge Demeyer, and Markus Borg.
\newblock Goal-oriented mutation testing with focal methods.
\newblock In {\em Proceedings of the 9th ACM SIGSOFT International Workshop on
  Automating TEST Case Design, Selection, and Evaluation}, A-TEST 2018, pages
  23--30, New York, NY, USA, 2018. ACM.

\end{thebibliography}
